# Obstructive Sleep Apnea Characterization: A Multimodal Cross-Recurrence-Based Approach for Investigating Atrial Fibrillation

Mantas Rinkevičius, Jesús Lázaro, Eduardo Gil, Pablo Laguna, Peter H. Charlton, Raquel Bailón, and Vaidotas Marozas

*Abstract* - **Obstructive sleep apnea (OSA) is believed to contribute significantly to atrial fibrillation (AF) development in certain patients. Recent studies indicate a rising risk of AF with increasing OSA severity. However, the commonly used apnea-hypopnea index in clinical practice may not adequately account for the potential cardiovascular risks associated with OSA. *(1) Objective*: to propose and explore a novel method for assessing OSA severity considering potential connection to cardiac arrhythmias. *(2) Method*: the approach utilizes cross-recurrence features to characterize OSA and AF by considering the relationships among oxygen desaturation, pulse arrival time, and heartbeat intervals. Multinomial logistic regression models were trained to predict four levels of OSA severity and four groups related to heart rhythm issues. The rank biserial correlation coefficient, $r_{rb}$, was used to estimate effect size for statistical analysis. The investigation was conducted using the MESA database, which includes polysomnography data from 2055 subjects. *(3) Results*: a derived cross-recurrence-based index showed a significant association with a higher OSA severity ($p < 0.01$) and the presence of AF ($p < 0.01$). Additionally, the proposed index had a significantly larger effect, $r_{rb}$, than the conventional apnea-hypopnea index in differentiating increasingly severe heart rhythm issue groups: 0.14 > 0.06, 0.33 > 0.10, and 0.41> 0.07. *(4) Significance*: the proposed method holds relevance as a supplementary diagnostic tool for assessing the authentic state of sleep apnea in clinical practice.**

Received Day Month Year; revised Day Month Year; accepted Day Month Year. Date of publication Day Month Year; date of current version Day Month Year. This work was supported by the project (22HLT01 QUMPHY) has received funding from the European Partnership on Metrology, co-financed from the European Union's Horizon Europe Research and Innovation Programme and by the Participating States. (Corresponding author: Mantas Rinkevičius.)

Mantas Rinkevičius is with the Biomedical Engineering Institute, Kaunas University of Technology, K. Barsausko str. 59, LT-51423 Kaunas, Lithuania (e-mail: mantas.rinkevicius@ktu.lt).

Jesús Lázaro, Eduardo Gil, Pablo Laguna, and Raquel Bailón are with the Biomedical Signal Interpretation and Computational Simulation (BSICoS) Group, Aragon Institute of Engineering Research (I3A), IIS Aragon, University of Zaragoza, 50009 Zaragoza, Spain, and also with the Biomedical Research Networking Center, 50018 Zaragoza, Spain (e-mail: jlazarop@unizar.es; edugilh@unizar.es; laguna@unizar.es; rbailon@unizar.es).

Peter H. Charlton is with the Department of Public Health and Primary Care, University of Cambridge, Cambridge CB2 1TN, U.K., and also with the Research Centre for Biomedical Engineering, University of London, London WC1E 7HU, U.K. (e-mail: pc657@medschl.cam.ac.uk).

Vaidotas Marozas is with the Faculty of Electrical and Electronics Engineering, Kaunas University of Technology, Studentu str. 50, LT-51368 Kaunas, Lithuania, and also with the Biomedical Engineering Institute, Kaunas University of Technology, K. Barsausko str. 59, LT-51423 Kaunas, Lithuania (e-mail: vaidotas.marozas@ktu.lt).

## I. INTRODUCTION

OBSTRUCTIVE sleep apnea (OSA) is a prevalent sleep-related breathing disorder characterized by recurrent upper airway obstruction, leading to diminished or absent breathing during sleep. Worldwide, it is estimated that over 1 billion people suffer from OSA with prevalence exceeding 50% in some countries [1]. It is reported that approximately 26% of adults aged 30-70 years experience symptoms of OSA [2]. However, 80–90% of OSA cases remain undiagnosed [3]. In addition, people with OSA often experience excessive daytime sleepiness [4] and loud snoring at night. Such patients are increasingly using healthcare services being at higher risk of type 2 diabetes [5], [6], [7], obesity [8], anxiety and mood disorders [9], [10], [11], and cardiovascular diseases [12], [13], [14], [15], [16], [17].

In clinical practice, the apnea-hypopnea index (AHI) is used as the gold standard for determining the presence and severity of OSA [18]. The AHI indicates the number of times that a patient stops breathing or experiences a significant reduction in airflow per hour of sleep time. According to the AHI, OSA is categorized into mild (5-15 events/hour), moderate (15-30 events/hour), and severe (> 30 events/hour) [19]. This index was introduced in 1983 [20] and is still currently used to describe OSA.

However, it has faced criticism for not capturing relevant clinical features and being an insufficient tool for predicting clinical outcomes [21], [22], [23], [24], [25], [26]. Critics argue that AHI thresholds lack validity for severity scoring, deeming severity categories arbitrary and potentially misleading for clinical decision-making [21], [27], [28]. Moreover, issues such as the appropriate definition of total sleep time in the denominator [21], [29], [30], the definition of criteria for hypopneas, and the consideration of event durations [21] have been raised. Currently, the primary role of the AHI as a diagnostic biomarker and severity indicator of clinically relevant OSA is declining [21]. We assert that the AHI alone is inadequate for assessing OSA severity, as it solely measures events per hour of sleep and fails to quantify other critical factors such as potential cardiovascular effects.

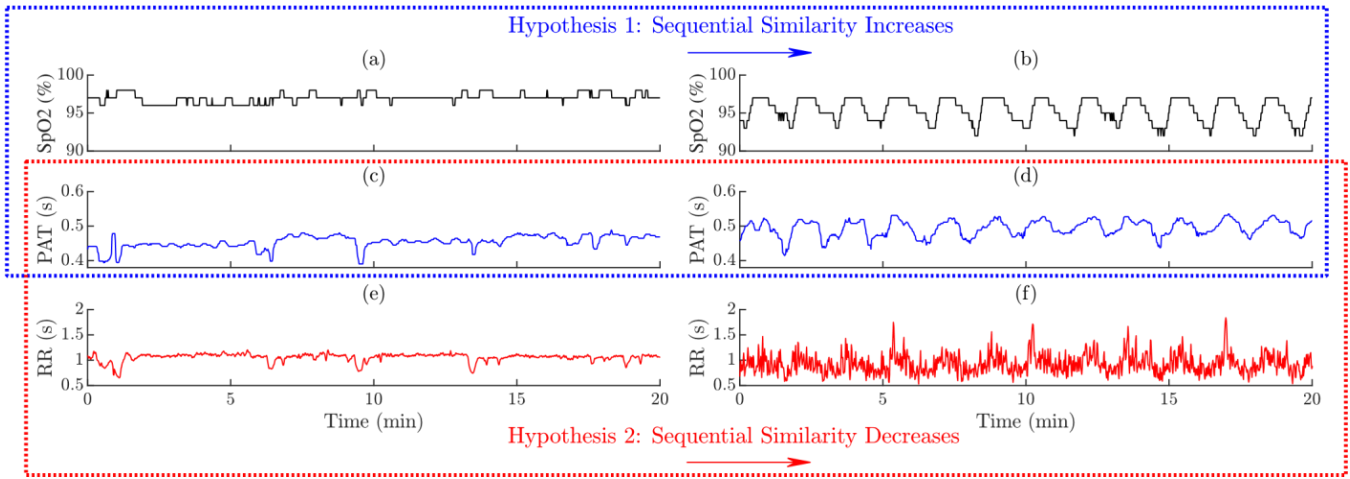

Fig. 1. Examples of time series estimated from PPG and ECG signals: SpO2 in (a) normal case and (b) during OSA and AF, blood pressure-correlated PAT in (c) normal case and (d) during OSA and AF, heart rate RR intervals in (e) normal case and (f) during OSA and AF.

Recent studies [12], [13], [14], [15], [16], [17], [31], [32], [33], [34] show that OSA and atrial fibrillation (AF) are potentially related. AF is a heart condition that leads to irregular and often abnormally fast atrial contractions when normal atrial systole no longer occurs. OSA is thought to be an important factor in the development of AF in certain sleep apnea patients [12], [31], [32]. However, it is currently unclear which patients with OSA are at increased risk of AF. Studies show that the risk of AF increases with increasing severity of OSA [35], [36]. Therefore, the presence of cardiac arrhythmias such as AF should be considered when assessing OSA in clinical practice.

We presume that potential arrhythmias in OSA may stem from abrupt heart rate changes associated with blood pressure fluctuations during apneic episodes [37]. During apnea, immediate cessation of breathing leads to a decrease in arterial blood oxygen saturation (SpO2) [38], [39], [40]. The sympathetic nervous system, triggered by stress, releases hormones, elevating both blood pressure and heart rate. Studies [38] indicate that apneic episodes with SpO2 desaturations are linked to decreased blood pressure-correlated pulse arrival time (PAT). With an increasing number of SpO2 desaturations and the severity of apnea, more simultaneous reductions in blood pressure-correlated PAT are expected. Consequently, we suggest that the sequential similarity between SpO2 and blood pressure fluctuations significantly increases with the severity of OSA due to larger desaturations (see Fig. 1 (a-d)). As oxygen deficiency induces changes not only in blood pressure, but also in heart rate, we hypothesize that cardiac arrhythmias cause a significant decrease in the sequential similarity between blood pressure fluctuations and heart rate time series in OSA patients (see Fig. 1 (c-f)).

To assess the sequential similarity between the mentioned time series, we employed cross-recurrence plot (CRP) analysis methods [41], [42], [43]. The CRP illustrates instances when the states of the first time series coincide with those of the second time series in a phase space trajectory. These methods are commonly used for analyzing the dynamical evolution similarity between different systems or investigating the time relationship of two similar systems [43]. CRPs provide useful information even for short intervals and non-stationary data, where other methods fail (e.g., Pearson correlation analysis) or could have different sensitivity due to non-stationarity (e.g., Dynamic Time Warping). For instance, the sensitivity of Dynamic Time Warping depends on various hyperparameters and factors such as the choice of distance metric, warp path constraints, the specific characteristics of the time series being compared, the algorithm and norms used, and the particular implementation [44], [45]. In addition, the CRP method offers a more straightforward interpretation than statistical approaches such as transfer entropy or mutual information, which are based on information theory.

This study introduces and explores a novel cross-recurrence properties-based method for characterizing OSA severity considering potential connection to cardiac arrhythmias. The proposed approach involves estimating cross-recurrence properties between SpO2 and blood pressure-correlated PAT time series, as well as between PAT and heart rate RR intervals.

The investigation of this study consists of the following stages: (i) cross-recurrence feature selection for models; (ii) analysis of cross-recurrence indexes estimated from selected features; (iii) comparison of obtained cross-recurrence indexes to the AHI; (iv) validation of implemented models.

## II. MATERIAL AND METHODS

### A. Dataset

In this study, the Multi-Ethnic Study of Atherosclerosis (MESA) dataset [46], [47] was used to implement and investigate the cross-recurrence properties-based approach to characterize OSA severity. The MESA data consists of 2055 patients aged 54–95 years old, totaling 16,300 hours of full overnight annotated polysomnography recordings. This sleep study included only subjects who had not used continuous positive airway pressure (CPAP), oxygen device, or received any other treatment for sleep apnea for more than a month before the study, or who reported using any of these treatments rarely, i.e., less than weekly.

Polysomnography signals were obtained at home using the Compumedics Somte monitoring system (Compumedics Ltd., Abbotsville, Australia). Electrocardiogram (ECG) and photoplethysmogram (PPG) signals with a sampling rate of 256 Hz were analyzed to evaluate cross-recurrence features between time series of SpO2 and PAT, and between PAT and RR intervals. Single-lead ECG signals were recorded from the chest using the Ag/AgCl electrode. Whereas PPG signals were obtained from the finger using the Nonin 8000 sensor.

The AHI values of MESA subjects were used to train and test the first cross-recurrence properties-based model for characterizing OSA severity. The subjects were categorized into four groups based on their AHI, which includes all observed apneas and hypopneas with ≥ 3% oxygen desaturation:
1) Normal / No Sleep Apnea (AHI < 5) - 414 subjects.
2) Mild Sleep Apnea (5 ≤ AHI < 15) - 643 subjects.
3) Moderate Sleep Apnea (15 ≤ AHI < 30) - 518 subjects.
4) Severe Sleep Apnea (AHI ≥ 30) - 480 subjects.

An apnea was defined as a complete or almost complete cessation of airflow, lasting at least 10 s, and usually associated with oxygen desaturations and/or arousal events. Whereas, a hypopnea was defined as a reduction in airflow (at least 50% of a baseline level), associated with oxygen desaturations and/or arousal events.

In addition, the MESA dataset includes labels of heart rhythm (HR) issues, thus, the subjects were categorized also into four other groups as follows:
1) No HR issues - 1801 subjects.
2) Abnormalities seen-not clinically significant - 194 subjects.
3) Urgent referral – HR (No AF) - 36 subjects. This group includes subjects with HR > 150 bpm or < 30 bpm for > 2 min, non sustained ventricular tachycardia, acute ST segment, oxygen saturation < 85% or other criteria, but excluding all AF cases.
4) Potential urgent - AF/flutter HR - 24 subjects. The group includes all cases of AF, regardless of rate, pre-existing diagnosis, duration or rhythm disturbance.

The mentioned HR issue groups were used to train and test the second cross-recurrence properties-based model for characterizing AF in sleep apnea patients. The distributions of OSA severity among HR issue groups are showed in Fig. 2.

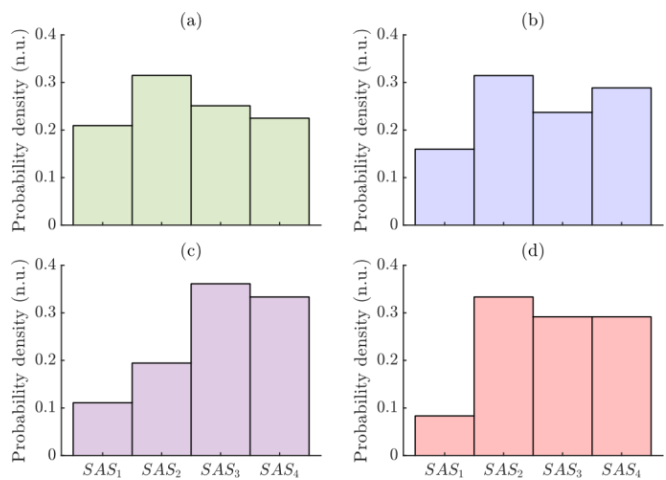

Fig. 2. The distributions of OSA severity ($SAS_1$ - No Sleep Apnea, $SAS_2$ - Mild Sleep Apnea, $SAS_3$ - Moderate Sleep Apnea, $SAS_4$ - Severe Sleep Apnea) among HR issue groups: (a) No HR issues, (b) Abnormalities seen-not clinically significant, (c) Urgent referral – HR (No AF), (d) Potential urgent - AF/flutter HR.

### B. The Proposed Structure of the Model Approach

The proposed cross-recurrence properties-based approach for characterizing OSA severity considering potential connection to AF consists of (see Fig. 3): (i) estimating time series of SpO2, PAT, and RR intervals; (ii) estimating cross-recurrence properties between SpO2 and PAT time series, and between PAT and RR time series; (iii) implementing models for characterizing OSA and AF in apnea patients; (iv) estimating cross-recurrence indexes from implemented hierarchical multinomial logistic regression models.

### C. ECG & PPG Signals Pre-processing

The ECG signals were processed using a zero-phase fourth-order Butterworth low-pass filter with a cut-off frequency of 25 Hz to mitigate high-frequency interferences. The baseline was removed by estimating it using a median filter with overlapping windows of 1 s duration and 0.5 s overlap, which was then interpolated and subtracted from the ECG signals [48]. The PPG signals were processed using a zero-phase fourth-order Butterworth band-pass filter with a pass-band of 0.4–6 Hz to reduce noise and improve the detectability of fiducial points.

### D. Assessing Time Series of SpO2 & PAT & RR

Oxygen desaturation-related SpO2 time series sampled at 1 Hz were provided in the MESA dataset, which were obtained from the finger using the Nonin 8000 sensor.

Blood pressure-correlated PAT time series were estimated as intervals between the ECG R peak and the subsequent PPG systolic peak. The PAT post-processing procedure involved correcting low-quality intervals and interpolating PAT sequences as described in [38].

RR time series representing heart-beats were obtained from filtered ECG signals using the R-DECO algorithm [49]. This algorithm has been found to perform well, with sensitivity of 99.6% and positive predictive value of 99.7% [49].

Fig. 1 shows examples of SpO2, PAT, and RR time series estimated from PPG and ECG signals in normal case and during OSA and AF episodes, resampled to the same sampling rate of 1 Hz.

### E. Estimation of Cross-Recurrence Properties

Visual analysis of CRPs can answer questions about the stationarity, periodicity, and randomness of time series, but it cannot accurately classify signals according to their complexity [43]. For this purpose, numerical cross-recurrence features have been proposed based on the density of black dots and the structures of diagonal, vertical, and horizontal lines in CRPs [50], [51]. For instance, black dots in the CRP represent the repetition of signal states. Diagonal lines represent episodes of quasi-periodicity in signals. The longer the diagonal line, the longer the signal was periodic. Vertical and horizontal lines represent episodes of permanency reflecting the presence of steady states in signals and indicating that these states change slowly in time. In terms of differences between these structures, the vertical lines represent co-occurrences of the $i^{th}$ signal states in the $j^{th}$ signal, whereas the horizontal lines do the opposite. It is stated that diagonal measures reveal pairwise similarity, as vertical and horizontal measures represent pairwise dissimilarity between time series [50].

In this study, we analyzed 10 cross-recurrence features [51], which were normalized within the range [0 1]. The normalization was performed by subtracting the minimum value of each feature and dividing by the difference between maximum and minimum values. These features include diagonal measures such as determinism, $DET$, mean diagonal length, $L$, maximal diagonal length, $L_{max}$, and entropy of diagonals, $ENTR$; and vertical and horizontal measures such as laminarity, $LAM_V$ and $LAM_H$, mean vertical and horizontal lengths known as trapping time, $TT_V$ and $TT_H$, and maximal vertical and horizontal lengths, $V_{max}$ and $H_{max}$. $DET$ measure is defined as the ratio of the number of dots forming diagonals to the total number of black dots. Whereas $LAM$ reflects the ratio of the number of dots forming vertical/horizontal lines to the total number of black dots.

The mentioned cross-recurrence features for OSA characterization were estimated from CRP diagrams, which were obtained from SpO2 and PAT time series:

$$CR_1(i,j) = \Theta(\epsilon - \|PAT_v(i) - SpO2_v(j)\|), \quad (1)$$

where $\Theta(\cdot)$ is the Heaviside unite step function, $\|\cdot\|$ is a norm, $\epsilon$ is a recurrence threshold distance between neighbouring states, while $i$ and $j$ are the time samples of $PAT_v$ and $SpO2_v$ time delay-embedded vectors, respectively. If the distance between two states of analyzed time series is less than this threshold, then we consider these states to be approximately repetitive in certain time instant ($i$, $j$).

In this work, we used the fixed recurrence rate method to define the threshold distance, $\epsilon$. It was defined as a fixed quantile of the recurrence distance matrix. The distances between two standardized vectors of states were calculated by using the Euclidean norm. The standardization procedure was performed by subtracting the mean and dividing by the standard deviation of time series.

To obtain time delay-embedded vectors, the embedding dimension, $m$, and the delay, $t$, parameters were used. The embedding dimension, $m$, defines a number of how many samples one signal state was composed. Whereas, the delay, $t$, is a sampling step, which was used to take samples from signals of time series.

The same cross-recurrence features for AF characterization were estimated from CRP diagrams, which were obtained from PAT and RR time series:

$$CR_2(i,j) = \Theta(\epsilon - \|RR_v(i) - PAT_v(j)\|), \quad (2)$$

where $i$ and $j$ are the time samples of $RR_v$ and $PAT_v$ time delay-embedded vectors, respectively.

Examples of CRP diagrams obtained from SpO2 and PAT time series, as well as from PAT and RR time series, are provided in Fig. 4.

### F. Hierarchical Multinomial Logistic Regression Models

The Hierarchical Multinomial Logistic Regression (HMLR) method was selected to implement models for characterizing OSA severity and AF, respectively. The most statistically important SpO2-PAT cross-recurrence features were selected as predictors of the HMLR Model 1 for assessing OSA severity. The features were ranked by using chi-square tests. To avoid high correlations among model parameters, the single most statistically significant feature was selected along with the most important ones from the diagonal, horizontal, and vertical structures, respectively. Similarly, the most statistically significant PAT-RR cross-recurrence features among HR issue groups were selected as predictors of the HMLR Model 2 to characterize AF in sleep apnea patients.

Further, two cross-recurrence indexes, $CRI^1$ and $CRI^2$, were estimated as log-odds expressions of the response probabilities related to the linear combination of the selected predictors from HMLR Models 1 and 2, respectively:

$$CRI^k = b_0^k + \sum_{i=1}^{i=4} b_i^k f_i^k, \quad (3)$$

where $k$ is the number of HMLR Model, $i$ is the indice of feature, and $b^k$ and $f^k$ are the coefficients and the selected features of HMLR Model $k^{th}$, respectively. The coefficients, $b^k$, were estimated for the models of the relative risk of having OSA ($SAS_2$, $SAS_3$, $SAS_4$ groups) / AF ($HRI_4$ group) versus not having OSA ($SAS_1$ group) / AF ($HRI_1$, $HRI_2$, $HRI_3$ groups) when performing training with 60% of the data and setting appropriate seed of the random number generator (see more details in Results Subsection A).

These indexes then were combined into a cross-recurrence index, $CRI$, for characterizing OSA severity considering potential connection to AF as follows:

$$CRI = \sqrt{CRI^1 CRI^2}, \quad (4)$$

where $CRI^1$ is the cross-recurrence index for OSA characterization, and $CRI^2$ is the cross-recurrence index for AF characterization among OSA patients.

To investigate which AF subjects could potentially have OSA-related AF, a single-term exponential model, $EM(AHI)$, was fitted on the cross-recurrence index, $CRI$, values sorted by the AHI from the AF group data:

$$EM(AHI) = ae^{(bAHI)}, \quad (5)$$

where AHI is the apnea-hypopnea index, and $a$ and $b$ are the model coefficients.

We used this exponential model to investigate whether the increasing number of sleep apnea episodes could lead to cardiac arrhythmias potentially associated with OSA.

### G. Statistical Analysis

The Anderson–Darling test found no Gaussian distribution in the analyzed cross-recurrence features and indexes. The nonparametric Mann-Whitney U test was used to test for statistical differences ($\alpha$ = 0.05) between cross-recurrence features and indexes for different apnea severity and HR issue groups. The following features and indexes were compared: (i) SpO2-PAT cross-recurrence features for apnea severity groups; (ii) PAT-RR cross-recurrence features for HR issue groups; (iii) cross-recurrence index $CRI^1$ distributions for apnea severity groups; (iv) cross-recurrence index $CRI^2$ distributions for HR issue groups; (v) cross-recurrence index $CRI$ distributions for apnea severity and HR issue groups; (vi) cross-recurrence index $CRI$ and AHI distributions for HR issue groups.

Additionally, the effect size was estimated using the rank biserial correlation coefficient, $r_{rb}$. For this purpose, median values for each subject were obtained and compared. The effect size is considered small when $r_{rb} < 0.30$, medium $r_{rb} \geq 0.30$, and large - $r_{rb} \geq 0.50$ [52].

### H. Performance Evaluation

In order to evaluate the performance of implemented HMLR models for characterizing OSA severity and AF in sleep apnea patients, the metrics for a particular class were estimated as sensitivity, $Se$, specificity, $Sp$, positive predictive value, $PPV$, negative predictive value, $NPV$, and accuracy, $Acc$. In addition, area under the ROC curve, $AUC$, was computed. Also, averaged metrics among classes were estimated.

To validate implemented HMLR models, 10-fold cross-validation was used. For reproducibility, the random number generator was set to the appropriate seed and Mersenne Twister algorithm.

### I. Data Mining for Class Balancing in HR Issue Groups

To address the imbalance between HR issue groups, the finite impulse response (FIR) antialiasing filter-resampling method was chosen for data mining and compared with the synthetic minority oversampling technique (SMOTE) [53].

The FIR Resampling method was used to resample the data so that each group contained approximately the same amount of data, while the total amount of data remained unchanged. The SMOTE method was used to additionally synthesize minority classes to reduce the imbalance between HR issue groups.

## III. RESULTS

### A. Parameter Settings

To estimate SpO2-PAT and PAT-RR cross-recurrence features, the average optimal embedding dimension across all subjects for SpO2, PAT, and RR time series was found to be $m$ = 5 by computing the amount of false nearest neighbours as a function of the embedding dimension (see Fig. 5 (a)). This approach involves embedding time series into a higher dimensional system and finding a dimension in which the neighbourhood in a lower dimension does not extend into a higher. The delay was set to $t = 1$, as it was observed that this parameter does not have a significant impact on the results as reported in [54]. The minimal length of diagonal, vertical, and horizontal lines was set to 2 to generate CRPs.

The fixed recurrence rate method was used to determine the size of neighbourhood, $\epsilon$. The optimal fixed recurrence rate of 7% was found by calculating the averaged rank biserial correlation coefficient, $r_{rb}$, across $CRI^1$ (Model 1) and $CRI^2$ (Model 2) distributions for apnea severity and HR issue groups, respectively (see Fig. 5 (b)). This was done by averaging the six $r_{rb}$ values obtained from the six possible pairs of sleep apnea severity and HR issue groups, when different fixed recurrences rates were used (see Fig. 5 (b) blue points for sleep apnea severity, and red points for HR issue groups). Similarly, the window size was set to 2% of the time series length, as the averaged proportion of $r_{rb}$ maximum locations (see Fig. 5 (c)). The overlap was selected to be half the window size.

.

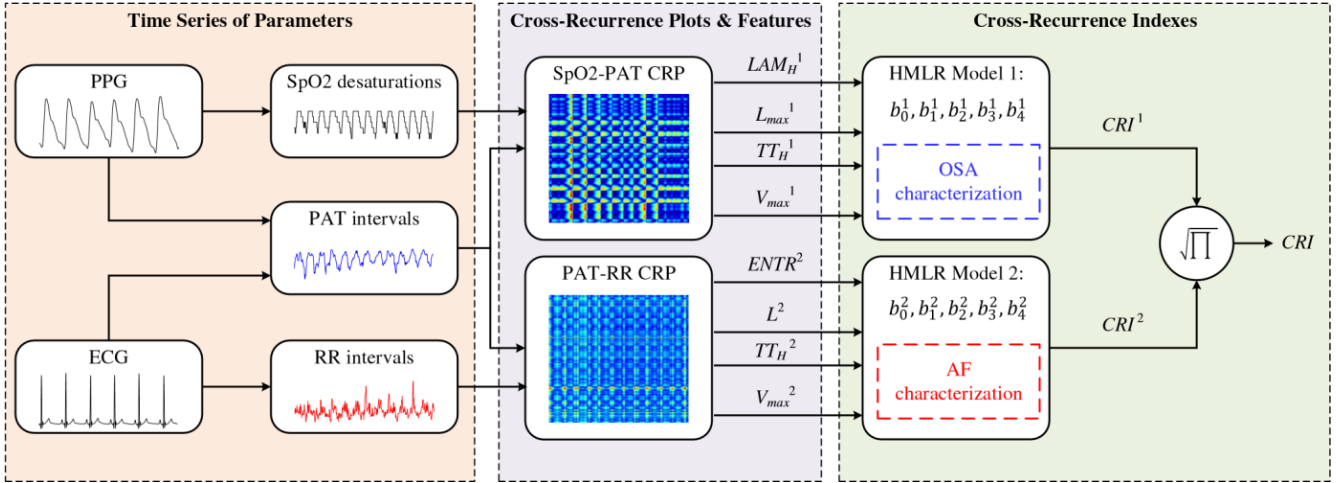

Fig. 3. The proposed structure of the cross-recurrence properties-based approach for estimating cross-recurrence indexes: CRI¹ for OSA, CRI² for AF, and CRI for OSA characterization considering potential connection to AF.

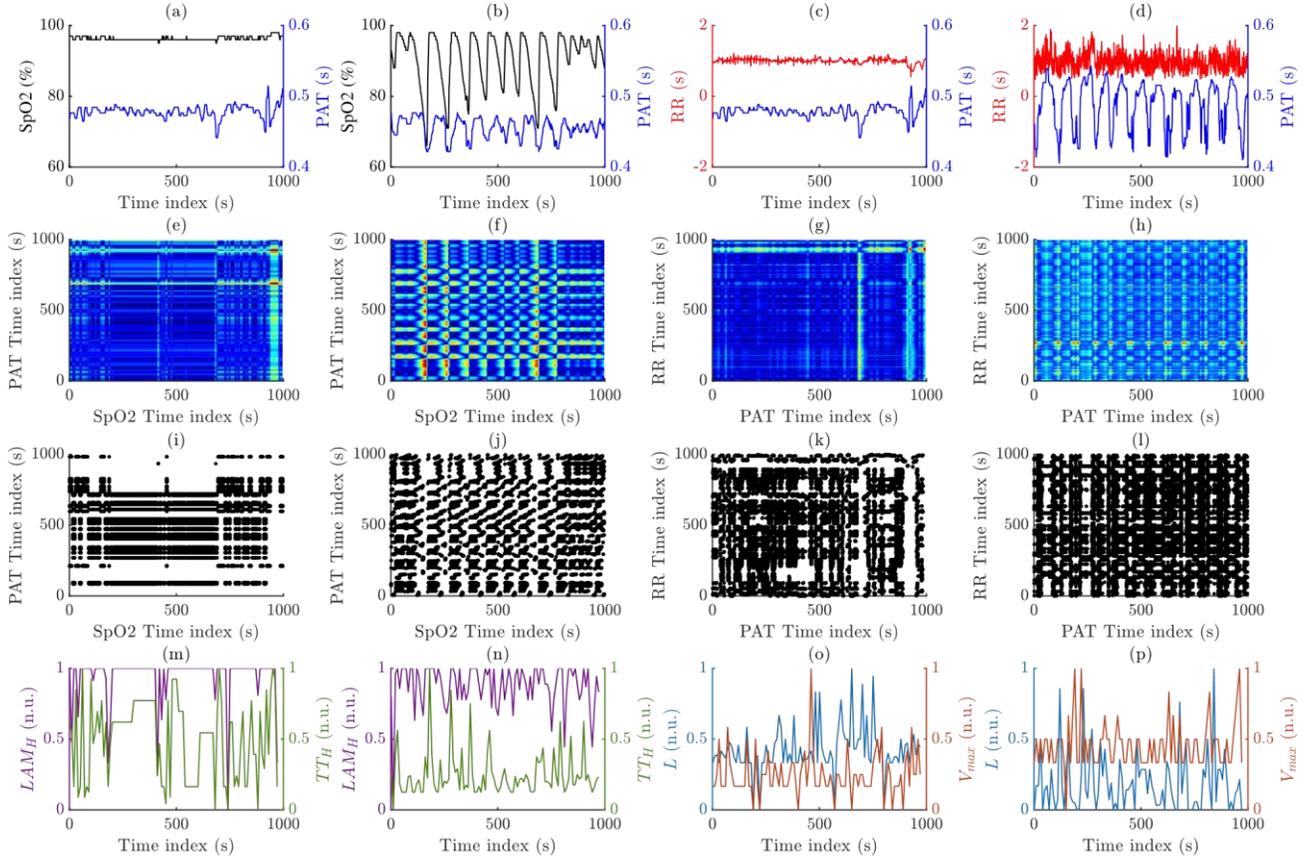

Fig. 4. Examples of CRP diagrams obtained from (a-b) SpO2 and PAT time series: unthresholded distance coded matrices – (e) normal case, (f) during OSA; thresholded matrices – (i) normal case, (j) during OSA; CRP exemplary features, laminarity, $LAM_H$, and trapping time, $TT_H$ – (m) normal case, (n) during OSA. Examples of CRP diagrams obtained from (c-d) PAT and RR time series: unthresholded distance coded matrices – (g) normal case, (h) during AF; thresholded matrices – (k) normal case, (l) during AF; CRP exemplary features, mean diagonal length, L, and maximal vertical length, $V_{max}$ – (o) normal case, (p) during AF.

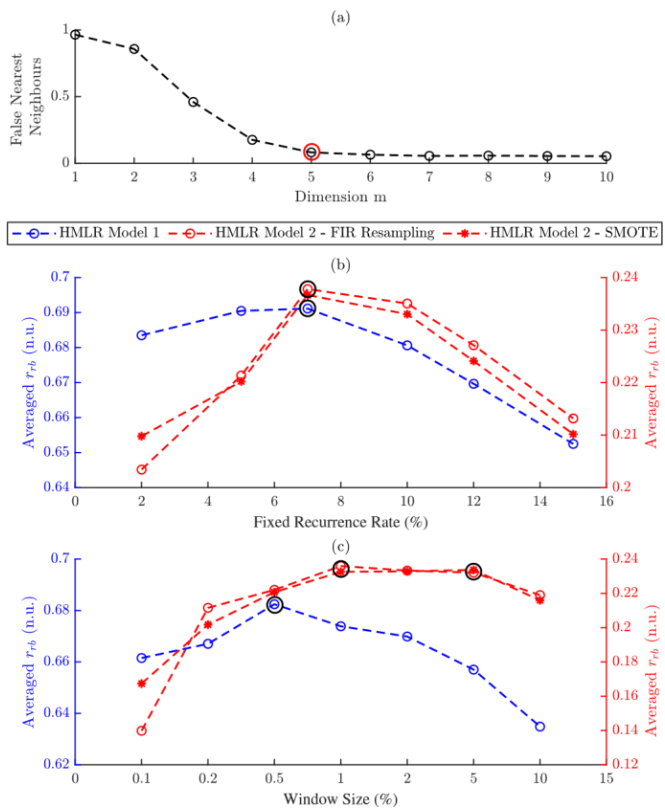

Fig. 5. Finding the optimal (a) embedding dimension by means of false nearest neighbours; (b) fixed recurrence rate; (c) window size by calculating the averaged rank biserial correlation coefficient, $r_{rb}$, across $CRI^1$ (Model 1 in blue) and $CRI^2$ (Model 2 in red) distributions for apnea severity and HR issue groups, respectively.

To ensure reproducibility in data splitting and mining, the Mersenne Twister random number generator was set to the $17^{th}$ seed. This seed was chosen as the minimum value that achieved a sensitivity of > 80% for identifying severe sleep apnea and AF when testing 40% of the data. The SMOTE method was used for data mining, with $k$ = 5 nearest neighbours considered during augmentation.

### B. Feature Selection for Models

The SpO2-PAT cross-recurrence features with the estimated rank biserial correlation $r_{rb}$ values for different sleep apnea severity groups are shown in Fig. 6. The $r_{rb}$ values are presented for three apnea severity classes comparing them to the severe sleep apnea severity group. The selected SpO2PAT cross-recurrence features for the HMLR Model 1 were $LAM_H$, $L_{max}$, $TT_H$, and $V_{max}$, with obtained coefficients of $b^1_0$ = -2.26, $b^1_1$ = 5.16, $b^1_2$ = -3.29, $b^1_3$ = 5.92, and $b^1_4$ = -0.97 (see Eq. 3), respectively. According to a decrease of horizontal lines-related features – $LAM_H$, $TT_H$, and $H_{max}$, a decrease in permanency between SpO2 and PAT time series is significantly associated with increasing severity of OSA (see Fig. 6: (h-j)). This means that as the number of apneic episodes increases, the frequency of co-occurrences of SpO2 steady states in PAT decreases.

The PAT-RR cross-recurrence features with the estimated rank biserial correlation $r_{rb}$ values for different HR issue groups are shown in Fig. 7. The $r_{rb}$ values are presented for three HR issue classes comparing them to the AF group. The selected PAT-RR cross-recurrence features for the HMLR Model 2 were $ENTR$, $L$, $TT_H$, and $V_{max}$, with obtained coefficients of $b^2_0$ = -2.47, $b^2_1$ = 20.95, $b^2_2$ = -21.52, $b^2_3$ = -1.87, and $b^2_4$ = 6.98 (see Eq. 3), respectively. According to a decrease of diagonal lines-related features – $DET$, $L$, $L_{max}$, and $ENTR$, a decrease in quasi-periodicity between PAT and RR time series is significantly associated with the presence of AF in OSA patients (see Fig. 7: (a-d)). Whereas a decrease of vertical lines-related features – $LAM_V$, $TT_V$, and $V_{max}$, shows that AF leads to the decreased frequency of co-occurrences of RR steady states in PAT (see Fig. 7: (e-g)).

### C. Analysis of Cross-Recurrence Indexes

The distributions of cross-recurrence indexes, $CRI^1$, $CRI^2$, and $CRI$, with the estimated rank biserial correlation $r_{rb}$ values for different sleep apnea severity and HR issue groups are provided in Fig. 8. The $r_{rb}$ values are presented for three respective classes comparing them to the fourth group.

Fig. 8 (a) reveals that a decrease in $CRI^1$ is significantly associated with increasing severity of OSA ($r_{rb}$ = 0.86, 0.81, 0.61, when $p$ < 0.001). This is likely due to $CRI^1$ being mostly related to a decrease in permanency between SpO2 and PAT time series.

Fig. 8 (b) shows that a decrease in $CRI^2$ is significantly associated with the presence of AF in OSA patients ($r_{rb}$ = 0.15, 0.40, 0.69, when $p$ < 0.001). This could be explained by the fact that $CRI^2$ is significantly related to a decrease in quasi-periodicity between PAT and RR time series during irregular rhythms.

Fig. 8 (c) and (d) indicate that a lower value of the new proposed cross-recurrence index, $CRI$, is significantly associated with a higher degree of OSA severity ($r_{rb}$ = 0.81, 0.75, 0.52, when $p$ < 0.01), especially with severe sleep apnea, and the presence of AF among OSA patients ($r_{rb}$ = 0.14, 0.33, 0.41, when $p$ < 0.01).

The coefficients of the single-term exponential model, $EM(AHI)$, were obtained as $a$ = 0.549 and $b$ = -0.009 with 95% confidence bounds of $a$ (0.468, 0.630) and $b$ (-0.015, -0.003). According to Fig. 8 (c) and (d) that a lower value of $CRI$ was significantly associated with a higher degree of OSA severity and AF, we assumed that $CRI$ decreases exponentially with increasing number of apnea episodes within AF subjects (see Fig. 9). Based on OSA characterization considering cardiac arrhythmias, Fig. 9 reveals that 15 out of 24 subjects are likely to have AF associated with OSA, while the remaining subjects are unlikely to have this association.

*D. Comparison to the AHI*

The proposed cross-recurrence indexes, $CRI^1$, $CRI^2$, and $CRI$, were compared to the AHI. The estimated rank biserial correlation $r_{rb}$ values of AHI distributions for different HR issue groups are provided in Fig. 8 (e). The $r_{rb}$ values are presented for three HR issue classes comparing them to the AF group. The new proposed cross-recurrence index, $CRI$, was found to be more sensitive to differentiate between HR issues, especially for AF ($r_{rb}$ - 0.14 > 0.06, 0.33 > 0.10, 0.41 > 0.07), which can be the advantage of the proposed method.

Fig. 10 shows the estimated rank biserial correlation $r_{rb}$ values among cross-recurrence indexes distributions for sleep apnea severity and HR issue groups. The $r_{rb}$ values among AHI distributions are provided for comparison (see Fig. 10 (b)). The cross-recurrence index, $CRI^1$, was found to have a larger effect size than the combined index, $CRI$, when differentiating among apnea severity groups (see Fig. 10 (a)). However, the combined index, $CRI$, had a significantly larger effect than the AHI when differentiating among HR issue groups (see Fig. 10 (b)). Anyway, $CRI^2$ outperforms $CRI$.

*E. Validation of Models*

The performance metrics of the HMLR Model 1 are provided in Table I. According to the prediction results, the severe sleep apnea group demonstrated the highest sensitivity of 76.24 ± 5.57%, positive and negative predictive values of 73.23 ± 8.61% and 92.62 ± 2.30%, respectively, accuracy of 87.84 ± 2.03%, and area under the ROC curve of 93.66 ± 1.76%. Whereas normal group demonstrated the highest specificity of 93.14 ± 1.51%.

The amount of data samples in each HR issue group after data mining methods is shown in Fig. 11.

The performance metrics of the HMLR Model 2 are presented in Table II. The AF group showed the highest sensitivity (FIR Resampling - 80.91 ± 4.24%, SMOTE – 81.69 ± 2.46%), positive predictive value (FIR Resampling - 64.91 ± 5.57%, SMOTE – 60.09 ± 3.34%), negative predictive value (FIR Resampling - 93.05 ± 1.99%, SMOTE – 95.08 ± 0.93%), accuracy (FIR Resampling - 84.43 ± 2.11%, SMOTE – 85.75 ± 1.54%), and area under the ROC curve (FIR Resampling - 89.69 ± 2.25%, SMOTE – 91.30 ± 1.38%). The highest specificity (FIR Resampling - 86.67 ± 2.48%, SMOTE – 89.99 ± 1.68%) was demonstrated by urgent referral HR group. The model did not show high estimates for other HR issue groups, which may be related to the clinically negligible degree of damage to the cardiovascular system.

Comparing the models, the HMLR Model 1 demonstrated the highest averaged accuracy of 80.46% ± 1.42% and area under the ROC curve of 85.26% ± 2.01% for characterizing OSA severity. For the HMLR Model 2, there are no significant differences between performance metrics by using FIR Resampling and SMOTE methods for data mining.

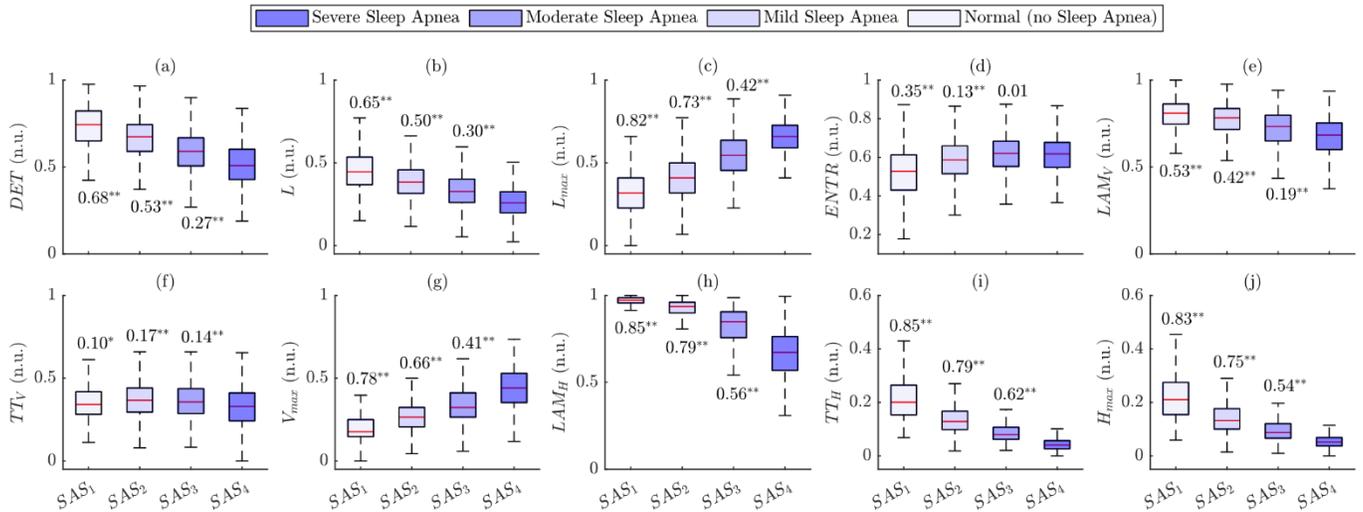

Fig. 6. The SpO2-PAT cross-recurrence features with the estimated rank biserial correlation $r_{rb}$ values for three apnea severity classes, $SAS_1$, $SAS_2$, and $SAS_3$, comparing them to the severe sleep apnea severity group, $SAS_4$ ($SAS_1$ - No Sleep Apnea, $SAS_2$ - Mild Sleep Apnea, $SAS_3$ - Moderate Sleep Apnea, $SAS_4$ - Severe Sleep Apnea): (a) determinism, DET, (b) mean diagonal length, L, (c) maximal diagonal length, $L_{max}$, (d) entropy of diagonals, ENTR, (e) laminarity of vertical lines, $LAM_V$, (f) trapping time of vertical lines, $TT_V$, (g) maximal vertical length, $V_{max}$, (h) laminarity of horizontal lines, $LAM_H$, (i) trapping time of horizontal lines, $TT_H$, (j) maximal horizontal length, $H_{max}$ (p< 0.05 is marked *, and p< 0.001 - **).

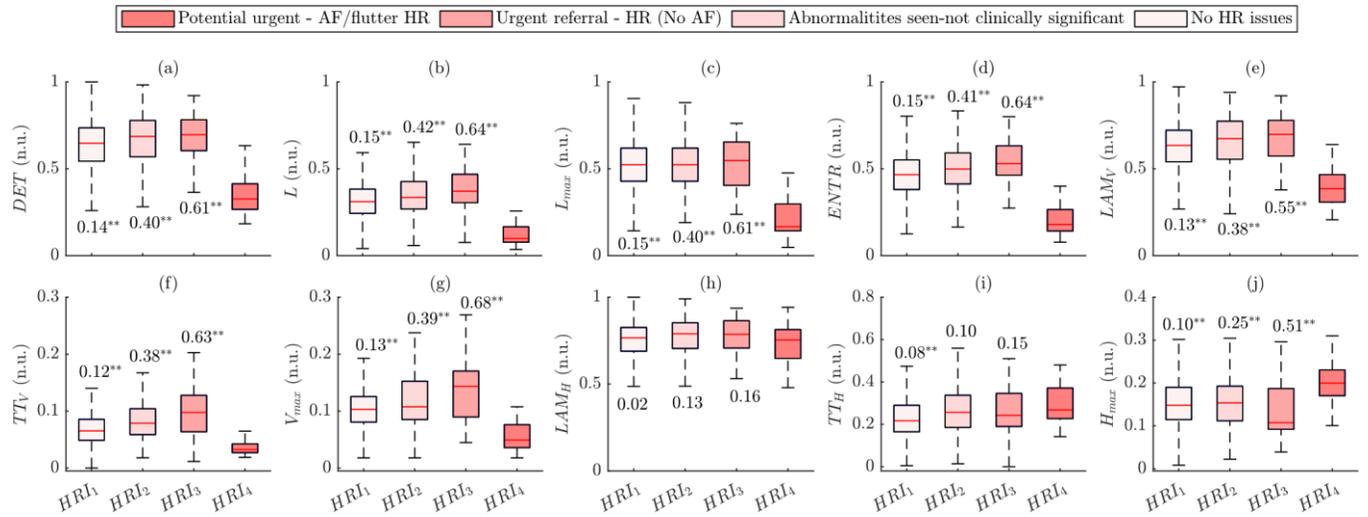

Fig. 7. The PAT-RR cross-recurrence features with the estimated rank biserial correlation $r_{rb}$ values for three HR issue classes, $HRI_1$, $HRI_2$, and $HRI_3$, comparing them to the AF group, $HRI_4$ ($HRI_1$ - No HR issues, $HRI_2$ - Abnormalities seen-not clinically significant, $HRI_3$ Urgent referral – HR (No AF), $HRI_4$ - Potential urgent - AF/flutter HR): (a) determinism, DET, (b) mean diagonal length, L, (c) maximal diagonal length, $L_{max}$, (d) entropy of diagonals, ENTR, (e) laminarity of vertical lines, $LAM_V$, (f) trapping time of vertical lines, $TT_V$, (g) maximal vertical length, $V_{max}$, (h) laminarity of horizontal lines, $LAM_H$, (i) trapping time of horizontal lines, $TT_H$, (j) maximal horizontal length, $H_{max}$ ($p < 0.05$ is marked *, and $p < 0.001$ - **).

TABLE I
THE PERFORMANCE METRICS OF THE HMLR MODEL 1 OBTAINED BY USING 10-FOLD CROSS-VALIDATION (MEAN ± STANDARD DEVIATION). THE HIGHEST METRICS AMONG GROUPS ARE BOLDED

| Apnea severity group | Se (%) | Sp (%) | PPV (%) | NPV (%) | Acc (%) | AUC (%) |
|---|---|---|---|---|---|---|
| Normal (no Sleep Apnea) | 57.50 ± 5.07 | **93.14 ± 1.51** | 67.78 ± 7.18 | 89.68 ± 1.61 | 85.94 ± 1.36 | 89.91 ± 2.31 |
| Mild Sleep Apnea | 65.55 ± 6.39 | 75.70 ± 3.84 | 55.14 ± 5.14 | 82.77 ± 4.00 | 72.36 ± 1.91 | 78.91 ± 2.27 |
| Moderate Sleep Apnea | 44.84 ± 8.43 | 86.34 ± 3.32 | 52.63 ± 7.39 | 82.25 ± 3.61 | 75.72 ± 2.37 | 78.55 ± 3.72 |
| Severe Sleep Apnea | **76.24 ± 5.57** | 91.49 ± 2.96 | **73.23 ± 8.61** | **92.62 ± 2.30** | **87.84 ± 2.03** | **93.66 ± 1.76** |
| Averaged metrics | 61.03 ± 2.70 | 86.67 ± 0.99 | 62.19 ± 3.16 | 86.83 ± 0.98 | 80.46 ± 1.42 | 85.26 ± 2.01 |

TABLE II
THE PERFORMANCE METRICS OF THE HMLR MODEL 2 OBTAINED BY USING 10-FOLD CROSS-VALIDATION (MEAN ± STANDARD DEVIATION). THE HIGHEST METRICS AMONG GROUPS ARE BOLDED

| HR issue group | Class Balancing Method | Se (%) | Sp (%) | PPV (%) | NPV (%) | Acc (%) | AUC (%) |
|---|---|---|---|---|---|---|---|
| No HR issues | FIR Resampling | 58.97 ± 10.74 | 77.24 ± 2.39 | 46.06 ± 4.74 | 85.01 ± 3.84 | 72.61 ± 2.98 | 78.26 ± 2.64 |
|  | SMOTE | 58.77 ± 3.72 | 70.19 ± 2.29 | 44.43 ± 3.55 | 80.75 ± 2.28 | 66.90 ± 2.60 | 71.66 ± 2.27 |
| Abnormalities seen-not clinically significant | FIR Resampling | 19.42 ± 5.12 | 84.04 ± 3.22 | 29.13 ± 8.05 | 75.77 ± 3.07 | 67.79 ± 2.72 | 55.65 ± 5.25 |
|  | SMOTE | 25.32 ± 5.08 | 80.85 ± 2.44 | 33.82 ± 5.48 | 73.63 ± 1.70 | 65.33 ± 2.35 | 60.15 ± 2.64 |
| Urgent referral – HR (No AF) | FIR Resampling | 41.61 ± 5.29 | 86.67 ± 2.48 | 51.04 ± 6.60 | 81.63 ± 3.12 | 75.43 ± 3.00 | 67.88 ± 4.89 |
|  | SMOTE | 28.38 ± 3.30 | 89.99 ± 1.68 | 46.89 ± 6.40 | 80.23 ± 1.82 | 75.41 ± 1.93 | 69.59 ± 2.55 |
| Potential urgent - AF/ flutter HR | FIR Resampling | 80.91 ± 4.24 | 85.60 ± 1.91 | **64.91 ± 5.57** | 93.05 ± 1.99 | 84.43 ± 2.11 | 89.69 ± 2.25 |
|  | SMOTE | **81.69 ± 2.46** | **86.76 ± 1.49** | 60.09 ± 3.34 | **95.08 ± 0.93** | **85.75 ± 1.54** | **91.30 ± 1.38** |
| Averaged metrics | FIR Resampling | 50.23 ± 3.13 | 83.39 ± 1.02 | 47.79 ± 2.90 | 83.86 ± 1.17 | 75.06 ± 1.56 | 72.87 ± 1.59 |
|  | SMOTE | 48.54 ± 2.61 | 81.95 ± 1.06 | 46.31 ± 3.50 | 82.42 ± 1.05 | 73.35 ± 1.56 | 73.17 ± 1.57 |

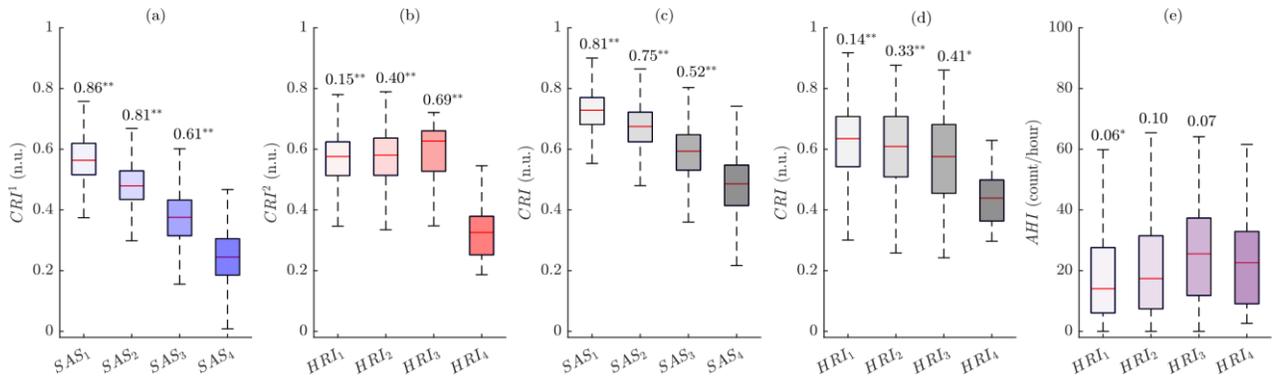

Fig. 8. The distributions of cross-recurrence indexes with the estimated rank biserial correlation $r_{rb}$ values: (a) $CRI^1$ for different sleep apnea severity groups ($SAS_1$ - No Sleep Apnea, $SAS_2$ - Mild Sleep Apnea, $SAS_3$ - Moderate Sleep Apnea, $SAS_4$ - Severe Sleep Apnea), (b) $CRI^2$ for different HR issue groups ($HRI_1$ - No HR

issues, HRI$_2$ - Abnormalities seen-not clinically significant, HRI$_3$ - Urgent referral – HR (No AF), HRI$_4$ - Potential urgent - AF/flutter HR), (c) CRI for different sleep apnea severity groups, (d) CRI for different HR issue groups, (e) AHI for different HR issue groups (p< 0.05 is marked *, and p< 0.001 - **).

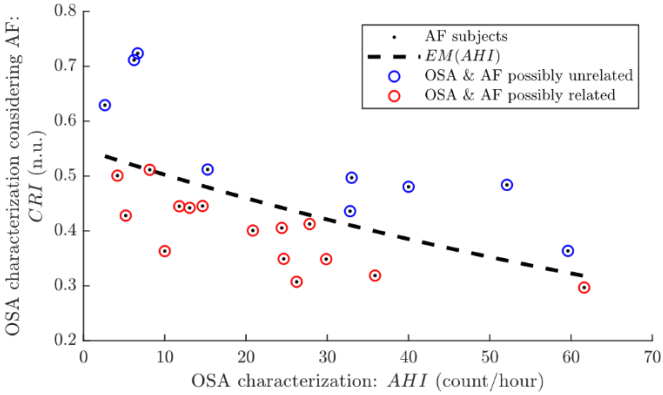

Fig. 9. The single-term exponential model, EM(AHI), fitted on the cross-recurrence index, CRI, values sorted by the AHI from the AF group data, with indicating subjects whose AF are possibly related to OSA (red circle points below the fitted curve) and subjects whose AF are possibly unrelated to OSA (blue circle points above the fitted curve).

## IV. DISCUSSION

*Main Findings*. The study aimed to propose a novel method for characterizing OSA severity considering potential connection to cardiac arrhythmias such as AF. The proposed approach was based on estimating cross-recurrence properties between SpO2 and blood pressure-correlated PAT time series, as well as between PAT and heart rate RR intervals. The method consisted of: (i) estimating time series of SpO2, PAT, and RR intervals; (ii) estimating cross-recurrence properties between SpO2 and PAT time series, and PAT and RR time series; (iii) implementing models for characterizing OSA and AF in apnea patients; (iv) estimating cross-recurrence indexes from implemented hierarchical multinomial logistic regression models. The key findings are as follows. First, a decrease in permanency between SpO2 and PAT time series is significantly associated with increasing severity of OSA. Second, a decrease in quasi-periodicity between PAT and heart rhythm interval RR time series is significantly associated with the presence of AF in OSA patients. Third, a lower value of the new proposed cross-recurrence index, *CRI*, is significantly associated with a higher degree of OSA severity, especially with severe sleep apnea, and the presence of AF in OSA patients. Fourth, the new proposed cross-recurrence index, *CRI*, is more sensitive to the presence of AF in OSA patients, and has a significantly larger effect than the AHI when differentiating among HR issue groups. Fifth, implemented models for characterizing OSA severity and AF demonstrated high performance metrics to identify severe sleep apnea and arrhythmias, respectively.

*Hypotheses Testing*. In this study, we tested two hypotheses. The first one stated that the sequential similarity between SpO2 and blood pressure fluctuations significantly increases with the severity of OSA due to larger desaturations.

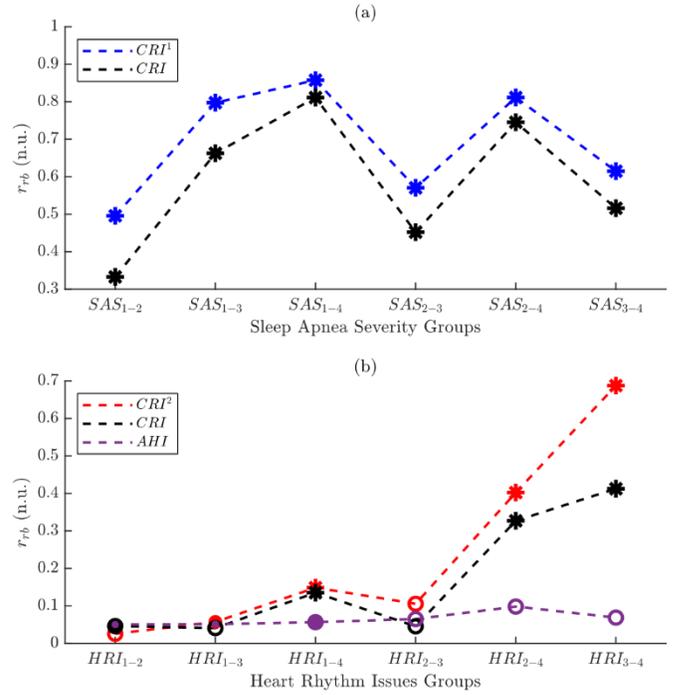

Fig. 10. The estimated rank biserial correlation $r_{rb}$ values among cross-recurrence indexes distributions for: (a) sleep apnea severity groups (SAS$_1$ - No Sleep Apnea, SAS$_2$ - Mild Sleep Apnea, SAS$_3$ Moderate Sleep Apnea, SAS$_4$ - Severe Sleep Apnea); (b) HR issue groups (HRI$_1$ - No HR issues, HRI$_2$ - Abnormalities seen-not clinically significant, HRI$_3$ - Urgent referral – HR (No AF), HRI$_4$ Potential urgent - AF/flutter HR). The subscripts indicate which groups were compared with each other. The $r_{rb}$ values among AHI distributions for HR issue groups are provided for comparison (p> 0.05 is marked °, p< 0.05 - •, and p< 0.01 - *).

The hypothesis was particularly confirmed by cross-recurrence features - maximal diagonal length, $L_{max}$, laminarity and trapping time of horizontal lines, $LAM_H$ and $TT_H$, respectively, and maximal horizontal length, $H_{max}$, estimated from SpO2 and blood pressure-correlated PAT time series. According to the results (see Fig. 6 (h-j)), with increasing severity of OSA, the sequential similarity between SpO2 and PAT significantly increases, because horizontal measures related to pairwise dissimilarity [50] decrease. Moreover, increasing diagonal-related feature, $L_{max}$, associated with pairwise similarity, also supports this assumption (see Fig. 6 (c)). However, other diagonal features, $DET$ and $L$, and SpO2-PAT vertical measures did not confirm the hypothesis. This suggests that the most appropriate estimate of diagonal-related similarity is $L_{max}$ representing the maximum quasi-periodic intervals of pairwise similarity in recurrence plots. On the other side, this may be due to the choice of the minimal diagonal length. Therefore, developing algorithms to determine such optimal values for recurrence analysis might be considered [55]. Moreover, it should be kept in mind that choosing too high a minimal diagonal length may miss important recurrent patterns.

We assume that a decrease in permanency between SpO2 and PAT was related to the increasing number of significant desaturation events, which led to the increasing number of simultaneous correlated PAT drops during apneic episodes via sympathetic activation [38]. Additionally, our studies have shown that SpO2-RR cross-recurrence features were not as successful in characterizing sleep apnea as SpO2-PAT analysis. This may indicate that the relation of SpO2 changes with PAT is more appropriate for apnea staging than the relation with RR intervals.

The second hypothesis was that cardiac arrhythmias cause a significant decrease in the sequential similarity between blood pressure fluctuations and heart rate time series in OSA patients. This was confirmed by decreasing diagonal measures ($DET$, $L$, $L_{max}$, and $ENTR$) related to pairwise similarity and estimated from PAT and heart-beat RR intervals (see Fig. 7 (a-d)). We presume that this decrease in quasi-periodicity between PAT and RR time series was related to cardiac rhythm disturbances during AF episodes. However, as with the first hypothesis, decreasing PAT-RR vertical measures did not support the second either. Based on our hypotheses (see Fig. 1) and obtained results (see Fig. 8 (a) and (b)), we could define $CRI^1$ as a measure of dissimilarity between SpO2 and PAT time series, whereas $CRI^2$ as a measure of similarity between PAT and RR time series.

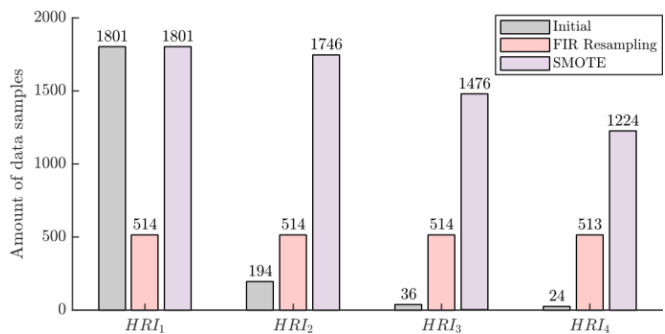

fig. 11. The amount of data samples in each HR issue group ($HRI_1$ - No HR issues, $HRI_2$ - Abnormalities seen-not clinically significant, $HRI_3$ - Urgent referral – HR (No AF), $HRI_4$ - Potential urgent - AF/flutter HR) after data mining methods.

*Validation Analysis*. We proposed two models to characterize OSA severity and AF, respectively. The first model was validated by dividing database patients into four groups based on AHI values. Whereas the second model was validated by dividing patients into four HR issue groups as provided in the MESA. Severe sleep apnea and AF classes demonstrated the highest performance metrics compared to other groups. This shows that the models perform better for characterizing the extremes. However, it should be noted that we used the AHI as the gold standard for validation, which has been criticized in the scientific literature [21], [22]. While the flawed AHI is currently used to confirm the OSA diagnosis, its limitations as the gold standard for OSA severity are well known [21]. Additionally, data mining methods were utilized for the second model due to the relatively low prevalence of AF among OSA patients, which may have influenced the validation results.

*Method Benefit*. Our study found that a lower value of the new proposed index, $CRI$, was significantly associated with a higher degree of OSA severity and the presence of AF among OSA patients. Moreover, our models showed high accuracy in identifying the stage of severe sleep apnea and AF. This is particularly important as our new proposed approach would allow further exploration of the relationships between sleep apnea and cardiovascular diseases in the future, helping to identify patients whose cardiac arrhythmias may be associated with OSA or individuals being at high risk of developing AF, and so potentially overcoming some of the limitations of the AHI. Study [56] concludes that OSA treatment with CPAP can reduce the risk of development and progression of AF. This kind of treatment leads to decreased sympathetic nervous system activity, arterial stiffness, blood pressure, and increased arterial baroreflex sensitivity [56]. As our proposed OSA characterization method considers AF, such an approach could be a clinically important biomarker to select OSA patients who would benefit from CPAP in term of AF risk reduction. In addition, the European Society of Cardiology recommends optimizing diagnosis and treatment of OSA to reduce AF recurrences and improve AF treatment results [57].

*Related Studies*. Recurrence plots have been used for analyzing sleep apnea in several studies [58], [59], [60]. However, our study is the first to propose the use of recurrence properties for multinomial OSA staging instead of apnea detection [61], [62]. In addition, our approach involves the integrated analysis of cardiac arrhythmias in sleep apnea patients. Unlike other studies [58], [59], which only used convolutional neural networks on recurrence plots, our study attempts a deeper interpretation of the estimated recurrence features at the level of apnea physiological analysis. The proposed approach is also distinctive in that it combines the analysis of different time series that can be obtained simultaneously. Such multimodal analysis may provide more crucial information on the physiological mechanisms of sleep apnea and its possible relationship to cardiac arrhythmias. Plausible arrhythmogenic mechanisms of sleep apnea were discussed in [63]. These mechanisms include respiratory effort during obstruction that results in the shortened atrial refractory period, intermittent hypoxemia and hypercapnia leading to sympathetic discharge, and intrathoracic pressure alteration leading to left atrial stretch and left ventricular afterload. Moreover, OSA is related with significant atrial remodeling characterized by atrial enlargement, reduction in voltage, widespread conduction abnormalities, and longer sinus node recovery [64]. These findings could also explain the association between OSA and AF diseases.

The study [65] showed that cardiac arrhythmias among OSA patients were related to age, male gender, body-mass index, and comorbidities such as hypertension, diabetes, dyslipidemia, and chronic obstructive pulmonary disease. The aim of our study was to propose a signal analysis-based method, which would be independent of demographic and anthropometric variables. However, monitoring the mentioned factors and investigating their associations may also be of interest to assess AF in patients with sleep apnea.

*Limitations*. Despite the promising results, this work has some limitations. Although studies [35], [36] show that AF is associated with increasing severity of OSA, it is unclear from the data used, which patients developed sleep apnea first rather than cardiac arrhythmias, and what is the origin of AF. It is important to note that cross-sectional studies such as ours have limitations in terms of establishing causality. Consequently, the findings should be interpreted as indicative of associations rather than causal relationships. The exponential model (see Fig. 9) was proposed to identify patients with potentially associated OSA and AF. By this study, we aim to propose a tool that could be potentially useful in future prospective studies dedicated to the establishment of causal relationships between OSA and AF.

Another limitation of the method is the interpretability of the proposed new index, *CRI*. Whereas $CRI^1$ and $CRI^2$ are related to measures of dissimilarity and similarity between the respective time series during OSA and AF, respectively, the physiological clarification of the combined index, *CRI*, is more complex.

*Future Work*. The proposed method could be simplified by using only PPG analysis to obtain the input time series. For instance, by replacing heart-beat RR intervals to pulse-to-pulse intervals and PAT sequences to PPG-derived morphological features correlated with blood pressure fluctuations. This simplification would enable the algorithms to be integrated into a wearable system and could be used not only for the diagnosis of OSA, but also for its prevention at home as well as its long-term monitoring. However, it should be noted that such simplification could lead to some uncertainty in the method.

## V. CONCLUSION

A novel cross-recurrence properties-based approach for characterizing OSA severity was proposed and explored. The derived cross-recurrence index showed a significant association with increasing OSA severity and the presence of AF among subjects with OSA. The proposed index was more sensitive than the conventional AHI in differentiating increasingly severe HR issues, especially for AF. The study demonstrates that the proposed method has the potential to be used as an alternative to the AHI and could be utilized as a supplementary tool to assess the authentic state of sleep apnea in clinical practice.

## ACKNOWLEDGMENT

*Author Contributions*. Methodology, processing, algorithms, analysis, figures, writing—first draft, M.R.; writing—review and editing, J.L.; writing—review and editing, E.G.; writing—review and editing, P.L.; writing—review and editing, P.H.C.; writing—review and editing, R.B.; supervision, conceptualization, writing—review and editing, V.M.

*Conflicts of Interest*. All authors have read the journals policy on disclosure of potential conflicts of interest and have none to declare. All authors have read the journals authorship agreement and the manuscript has been reviewed and approved by all authors.